\pdfoutput = 1
\documentclass[epj]{svjour}

\usepackage{graphics}
\usepackage{epsfig}
\usepackage{hyperref}
\usepackage{amssymb}
\usepackage{amsmath}
\usepackage{amsfonts}
\usepackage{bm}
\usepackage{braket}
\usepackage{graphicx}
\usepackage{bbold}
\usepackage{xcolor}
\newcommand{\nn}{\nonumber}
\newcommand{\ud}{{\textrm{d}}}

\begin{document}
\title{Dynamics of Disordered Quantum Systems Using Flow Equations}

%An introduction to flow equations for the dynamics of disordered systems}
\author{S. J. Thomson\inst{1}
\thanks{\emph{Present address:} Centre de Physique Th\'{e}orique, CNRS, Institut Polytechnique de Paris, Route de Saclay, F-91128 Palaiseau, France. \emph{E-mail:}  steven.thomson@polytechnique.edu}%
 \and M. Schir\'o\inst{1,2}
}                     % Do not remove
\institute{Institut de Physique Th\'{e}orique, Universit\'{e} Paris Saclay, CNRS, CEA, F-91191 Gif-sur-Yvette, France  \and JEIP, USR 3573 CNRS, Coll\'ege de France, PSL Research University, 11 Place Marcelin Berthelot, 75321 Paris Cedex 05, France}
\date{Received: date / Revised version: \today}
\abstract{
In this manuscript, we show how flow equation methods can be used to study localisation in disordered quantum systems, and particularly how to use this approach to obtain the non-equilibrium dynamical evolution of observables. We review the formalism, based on continuous unitary transforms, and apply it to a non-interacting yet non trivial one dimensional disordered quantum systems,  the power-law random banded matrix model whose dynamics is studied across the localisation-delocalisation transition. We show how this method can be used to compute quench dynamics of simple observables, demonstrate how this formalism provides a natural framework to understand operator spreading and show how to construct complex objects such as correlation functions. We also discuss how the method may be applied to interacting quantum systems, and end with an outlook on unsolved problems and ways in which the method can be further developed in the future. Our goal is to motivate further adoption of the flow equation method, and to equip and encourage others to build on this technique as a means to study localisation phenomena in disordered quantum systems.
\PACS{
      {72.15.Rn}{Localization effects (Anderson or weak localization)}   \and
      {05.10.Cc}{Renormalization group methods}
     } % end of PACS codes
} %end of abstract
\maketitle
\section{Introduction}
\label{intro}

The addition of disorder into a quantum system can lead to all manner of rich new physics far from thermal equilibrium. From the paradigmatic phenomenon of Anderson localisation \cite{Anderson58,Evers+08,Gornyi+05} to the more recently-discovered many-body localisation \cite{Basko+06,Oganesyan+07,Pal+10,Imbrie16a,Imbrie+16b,Altman+15,Nandkishore+15,Alet+18,AbaninRMP}, and even through to vastly different physical systems such as quantum glasses \cite{Fisher+89,Cugliandolo+98,Cugliandolo+99,Cugliandolo+01,Biroli+01,Biroli+02}, disorder can have dramatic effects. In many-body localised (MBL) systems, the interplay of disorder and quantum fluctuations can protect quantum order in regimes where it would not otherwise exist \cite{Huse+13}, and in driven systems it can prevent heating to infinite temperature~\cite{PonteEtAlPR15} and instead stabilise unusual forms of matter that do not exist in thermal equilibrium \cite{Wilczek12,KhemaniEtAlPRL16,Else+16,Zhang+17,Choi+17}. 

Despite the substantial progress made, there are still many outstanding questions as to the role of disorder even in ostensibly simple physical systems, such as models of non-interacting fermions - see Ref. \cite{Evers+08} for a longer discussion. In particular, with the advent of highly controllable experiments in degenerate quantum gases \cite{Billy+08,Kondov+15,Choi+16,Choi+17} and trapped ion systems \cite{Smith+16,Zhang+17}, it is now possible to experimentally probe the \emph{dynamics} of disordered quantum matter in a way that has until now been impossible. In light of this development, it is important to develop our theoretical methods for the study and simulation of the non-equilibrium dynamics of these complex systems. Here, we will study the metal-insulator transition in a disordered system of non-interacting fermions with long-range, power-law-decaying hopping.

Our motivations here are threefold: i) to study the dynamics of an interesting, exactly solvable yet non-trivial disordered system with a tractable metal-insulator transition; ii) to lay the groundwork for future studies of this model in the presence of interactions, which may exhibit an interesting and unusual many-body localised phase; iii) to provide a detailed description of the flow equation technique recently adopted by several research groups to study (many-body) localisation in quantum systems isolated from their environments \cite{Quito+16,Pekker+17,Savitz+17,Thomson+18,Kelly+19,Yu+19,Savitz+19}, and in particular outline how to use it to study non-equilibrium dynamics. 

The layout of this manuscript is as follows: in Sec. 2, we will describe the technique of flow equations (or `continuous unitary transforms') \cite{Kehrein07} and show how they can be used to compute the non-equilibrium dynamics of observables. In Sec. 3, we shall apply the method to the Anderson model of non-interacting spinless fermions on a lattice, and discuss how different choices of generators for the unitary transform can lead to unitary flows with very different properties. In Sec. 4, we specify to the Wegner generator widely used in the literative and study a system of non-interacting fermions with power-law-decaying hopping, known as the Power-Law Random Banded Matrix (PRBM) model \cite{Levitov90,Mirlin+96,Levitov99}. We first demonstrate how the technique works by computing static properties of the Hamiltonian, namely energy eigenvalues and level spacing distributions. We shall then go on to compute quench dynamics and show how flow equations can be used to obtain the dynamics of local operators, global quantities and correlation functions, as well as provide some insight into the phenomenon of `operator spreading' \cite{Nahum+18,Khemani+18,Gopalakrishnan+18}. We then conclude this manuscript with a discussion of interesting open problems and challenges for both the development of the flow equation method and for future dynamical studies of disordered systems, including an outlook for the PRBM model.

\section{The Flow Equation Method}
\label{sec:method}
The central idea behind the flow equation method is to use a continuous unitary transform to diagonalise the Hamtiltonian. Once diagonalised, we can compute the time evolution of any observable by transforming the observable into the diagonal basis, time-evolving with respect to the diagonal Hamiltonian, and then transforming back to the original, physical basis. Before getting to that, however, let us first discuss how to diagonalise a Hamiltonian using this procedure.

\subsection{Diagonalising the Hamiltonian}

The use of unitary transforms to diagonalise Hamiltonians is common practice in many fields and examples are abundant, from non-interacting quantum particles on a lattice which can be diagonalised with a Fourier transform to interacting systems which can be perturbatively diagonalised. The Schrieffer-Wolff transformation, for example, falls into the latter category. Consider a Hamiltonian $\mathcal{H}$ with diagonal component $\mathcal{H}_0$ and off-diagonal component $V$:
\begin{align}
\mathcal{H} &= \mathcal{H}_0 + V, \\
\tilde{\mathcal{H}} &= \textrm{e}^{S} \mathcal{H} \textrm{e}^{-S} ,\\
& = \mathcal{H} + [S,\mathcal{H}] + \frac12 [S,[S,H]]+...,
\end{align}
and by choosing the generator of the transform $S$ such that $[S,\mathcal{H}_0]= -V$, one is left with a diagonal Hamiltonian given (to leading order) by:
\begin{align}
\tilde{\mathcal{H}} & \approx  \mathcal{H} + [S,V].
\end{align}
Here, the accuracy of the transform is controlled by how many of the infinite series of nested commutators we evaluate. Instead, we can imagine making many infinitely small transforms, such that the error at each step is formally infinitesimal. This series of infinitesimal transforms can be written as a single \emph{continuous unitary transform}. We parametrise the transform in terms of a fictitious `flow time' $l$, which runs from $l=0$ (the microscopic model) to $l = \infty$ (the diagonalised Hamiltonian). With an appropriate choice of generator $\eta(l)$ (to be discussed later), each infinitesimal transform takes the form:
\begin{align}
\mathcal{H}(l + \ud l) &= \textrm{e}^{\eta(l) \ud l} \mathcal{H}(l) \textrm{e}^{-\eta(l) \ud l}, \\
&= \mathcal{H}(l) + \ud l \phantom{.} [\eta(l),\mathcal{H}(l)] .
\end{align}
This allows us to recast the diagonalisation process in terms of a first-order ordinary differential equation:
\begin{align}\label{eqn:flow}
\frac{\ud \mathcal{H}}{\ud l} = [\eta(l),\mathcal{H}(l)].
\end{align}
This `flow equation' for the Hamiltonian allows us to interpret the transform in a way similar to renormalisation group flow. It was originally introduced to physicists by by Glazek and Wilson \cite{Glazek+93,Glazek+94} and independently proposed by Wegner \cite{Wegner94}, though it had been previously used by mathematicians under the names `double bracket flow' \cite{Brocket91} and `isospectral flow' \cite{Chu+90,Chu94}. A key difference between this and typical renormalisation group (RG) methods is that while RG iteratively eliminates irrelevant degrees of freedom, the flow equation method instead keeps all information throughout the flow. For non-interacting systems, this can be accomplished exactly. For interacting systems, which we shall discuss only briefly at the end of this manuscript, the situation is typically more complicated and additional approximations are necessary in order to obtain an analytic set of flow equations~\cite{Wegner06,Kehrein07,Thomson+18}. The flow equation method is equivalent to a single unitary transform $U(l)$ given by:
\begin{align}
U(l) = \mathcal{T}_l \exp \left[ \int_0^l \ud l' \eta(l') \right] \label{eq.U},
\end{align}
where the $\mathcal{T}_l$ signifies time-ordering with respect to the fictitious `flow time' $l$.

\subsubsection{Choice of the Flow Generator}
\label{sec.generator}

The unitary transform used to diagonalise the Hamiltonian is not unique, and different transforms will accomplish the goal in different ways. In this formalism, the determining factor is the choice of the generator $\eta(l)$. We must choose this generator such that the $l \to \infty$ Hamiltonian becomes diagonal. A common choice is the Wegner generator \cite{Wegner94,Kehrein07} (or `canonical generator') given by
\begin{align}
\eta = [\mathcal{H}_0,V],
\end{align}
where $\mathcal{H}_0$ is the diagonal part of the Hamiltonian in a given basis and $V$ contains the off-diagonal elements. Using this choice one can show that the off diagonal components of the Hamiltonian monotonically decay under the flow at long flow times~\cite{Monthus16}, 
which guarantees that the Hamiltonian will eventually become diagonal under this unitary transform. This generator is a good choice because it is numerically robust. The main downside of this generator is that, as we shall see, it does not preserve the sparsity of the Hamiltonian: during the initial stages of the flow, it will typically generate new off-diagonal couplings. In general, any possible couplings allowed by the symmetry of the model will be generated during the flow, For non-interacting systems, these terms will be long range hopping terms, while for interacting systems they will include new higher-order interaction terms. Any newly-generated off-diagonal couplings will ultimately decay to zero under the transform, but they must be accounted for during the flow in order to obtain the correct answer, else the transform will no longer be strictly unitary.
An alternative choice which preserves the sparsity of the Hamiltonian is given by the Toda generator \cite{Monthus16}:
\begin{align}
\eta = \sum_{ij} \textrm{sgn}(i-j) \mathcal{H}_{ij}.
\end{align}
This generator does not guarantee the monotonous decay of the off diagonal elements of the Hamiltonian, as it allows them to temporarily increases during the flow in order to maintain the sparsity of the model. It can be understood as a continuous limit of the QR algorithm for diagonalising matrices and a detailed description of its properties can be found in Ref. \cite{Monthus16}, however we will not go into the details here. In fact, we will not dwell on this generator as it is numerically difficult to implement. The central reason is that it rearranges the eigenvalues of the Hamiltonian into ascending order: this continual `reshuffling' of eigenvalues throughout the flow leads to a less well-controlled flow than the Wegner generator. Example flows of Toda and Wegner are shown in Fig. 1 for the simple case of a one-dimensional Anderson insulator, illustrating the differences between the flow generated by these two different unitary transforms.

We have only discussed two possible popular choices of generator here, however there are of course many more, as can be found in Refs. \cite{Monthus16,Savitz+17}. Constructing new generators better tailored to quantum many-body systems is an open problem, and there is significant scope for future work on this topic: in particular, a generator combining the convergence properties of Wegner flow with the sparsity-preserving properties of Toda flow would be highly desirable for studies of many-body localisation.

\subsection{Computing Observables}
The result of the flow-equation gives us direct access to the eigenvalues of the problem as well as in principle to the unitary $U(\infty)$ that diagonalises the Hamiltonian, which can be constructed from the expression given in Eq. \ref{eq.U}. From this unitary operator, one can also compute the eigenstates of the Hamiltonian, though for all but the simplest systems it can be difficult to explicitly calculate $U(\infty)$ due to the presence of the time-ordering operator $\mathcal{T}_l$. In practice, rather than simply computing eigenvalues we might want to compute the average of a given observable $A$ over a certain state (or density matrix) of the system. For an equilibrium density matrix $\rho=e^{-\beta H}/Z$, for example, we want
$$
\langle A\rangle=\mbox{Tr}\rho A=\mbox{Tr}\rho(l) A(l),
$$
where $\rho(l)=e^{-\beta H(l)}/Z$ is the flow-evolved density matrix, while $A(l)=U(l)AU^{\dagger}(l)$ is the flow-evolution of the operator $A$ which satisfies an analogous flow evolution
\begin{equation}
\frac{\ud A}{\ud l} = [\eta(l),A(l)].
\end{equation}
To access observables within this scheme one has to perform a flow evolution up to $l=\infty$ of both Hamiltonian and observables and then evaluate the trace in the basis where the Hamiltonian is diagonal. A similar strategy can be followed to access real-time evolution, as we are going to discuss below.

\subsection{Real-Time Dynamics}
We can use the flow equation formalism to compute the dynamics of any given observable, as in Refs. \cite{Moeckel+08,Hackl+08,hackl+09,Thomson+18}. Consider the case where we have a complicated, off-diagonal microscopic Hamiltonian $\mathcal{H}$ which generates unitary time evolution. 
The evolution of a generic operator $A_t=e^{iHt}Ae^{-iHt}$ can be written, by inserting the unitary $U(l)$ at scale $l$, as
\begin{eqnarray}
A_t &=&  e^{iHt} Ae^{-iHt}\\
=&&U(l)e^{i H(l)t} A(l) e^{-iH(l)t} U^{\dagger}(l) ,
\end{eqnarray}
where $A(l)=U(l)AU^{\dagger}(l)$ is the flow-evolution of the operator $A$ while $H(l)=U(l)HU^{\dagger}(l)$ is the flow of the Hamiltonian. The idea is then to run the flow up to $l=\infty$ where the Hamiltonian becomes diagonal, while tracking the flow of $A$, then perform the time evolution up to time $t$, which is now trivial in the diagonal basis, and then reverse the flow backward to come back to the original basis where the problem is formulated.  In practice, we first evolve the Hamiltonian according to the Wegner flow~(\ref{eqn:flow}) and the observable $A$ according to the equation (forward evolution)
\begin{align}
\frac{\ud A}{\ud l} = [\eta(l),A(l)],
\end{align}
with the initial condition $A(l=0) = A$. Once we have time-evolved the observable $A(\infty)$ with Hamiltonian $H(\infty)$, we compute the transform back into the original basis (backwards transform) using the same equation above, but now integrating from $l=\infty \to l=0$ with the new initial condition of $A(\l = \infty) = \tilde{A}(t)$ (where the tilde signifies the time-evolved operator is in the $l=\infty$ basis).

\section{Non-interacting fermions: Anderson Model}
\label{sec:anderson}

As a concrete example, let us consider the case of non-interacting fermions in one dimension moving in a disordered potential.
\begin{align}
\mathcal{H} = \sum_{i} J c^{\dagger}_i c_{i+1} + \sum_i h_i n_i,
\end{align}
where $h_i$ are uncorrelated random variables drawn from a box distribution of width $[-W,W]$. This is the Anderson model which can be solved exactly using, e.g. exact diagonalisation or transfer matrix methods. In one dimension, the Anderson model is always localised for any finite concentration of disorder, however with a different choice of $J_{ij}$, this Hamiltonian can also describe a physically richer example as we will discuss more in detail below.

The very first thing we find when we try to compute the flow equations using Eq.~\ref{eqn:flow} is that nearest-neighbour hoppings are generated; if we include these in the Hamiltonian from the beginning and compute the flow equations again, we will find that next-nearest neighbour hoppings are generated, and so on. In fact, if we wish to obtain a closed set of flow equations, we must allow for all possible long-range hopping terms. For the \emph{running Hamiltonian} $\mathcal{H}(l)$, we make the following ansatz:
\begin{align}
\mathcal{H}(l) = \sum_{ij} J_{ij}(l) c^{\dagger}_i c_j + \sum_i h_i(l) n_i,
\end{align}
with initial conditions $J_{ij}(l=0) = \delta_{j,i \pm1} J_0$ encoding the nearest-neighbour hopping of the microscopic model. The important step here is that all coefficients in the Hamiltonian are now explicit functions of the fictitious `flow time' $l$, and that the sum in the kinetic term runs over \emph{all} combinations of indices. As discussed in Sec. \ref{sec.generator}, this is because under Wegner flow, all possible long-range couplings with be generated at intermediate stages during the flow, though they will eventually be eliminated in the $l \to \infty$ limit, leaving us with the final fixed-point Hamiltonian of the form $\tilde{\mathcal{H}}(l \to \infty) = \sum_i \tilde{h}_i n_i$, where the $\tilde{h}_i$ are the eigenvalues of the problem. For the non-interacting system, this ansatz is in fact exact: these long-range hopping terms are the only new terms that the transform can generate. For interacting systems, this is not the case, and wholly new high-order terms can and typically will be generated under the flow: these require special care to handle (see, e.g., Refs. \cite{Kehrein07,Thomson+18,Kelly+19}) and we shall not discuss them here.

\subsection{The generator}
The Wegner generator takes the following form\:
\begin{align}
\eta(l) &= [\mathcal{H}_0(l),V(l)] = \left[ \sum_i h_i(l) n_i , \sum_{jk} J_{jk}(l) c^{\dagger}_j c_k \right].
\end{align}
The basic commutation relation we will use throughout is given by:
\begin{align}
\left[ c^{\dagger}_{\alpha} c_{\beta} , c^{\dagger}_{\gamma} c_{\delta} \right]  &= c^{\dagger}_{\alpha}  \{ c_{\beta} , c^{\dagger}_{\gamma} \} c_{\delta} - c^{\dagger}_{\gamma} \{ c_{\delta}, c^{\dagger}_{\alpha} \} c_{\beta} \\
&= \delta_{\beta \gamma} c^{\dagger}_{\alpha} c_{\delta} - \delta_{\delta \alpha} c^{\dagger}_{\gamma} c_{\beta}.
\end{align}
Note that the final result of this commutator is the same for both fermions and bosons. Evaluating the generator using this expression, we get:
\begin{align}
\eta(l) &=  \sum_{ij} J_{ij}(l) (h_{i}(l) -h_{j}(l) ) c^{\dagger}_i c_{j}.
\end{align}
We see that the generator is an anti-Hermitean operator. For clarity, in the following we will suppress the explicit dependence of the running constants on flow time $l$.
\begin{figure}
\resizebox{0.975\linewidth}{!}{\includegraphics{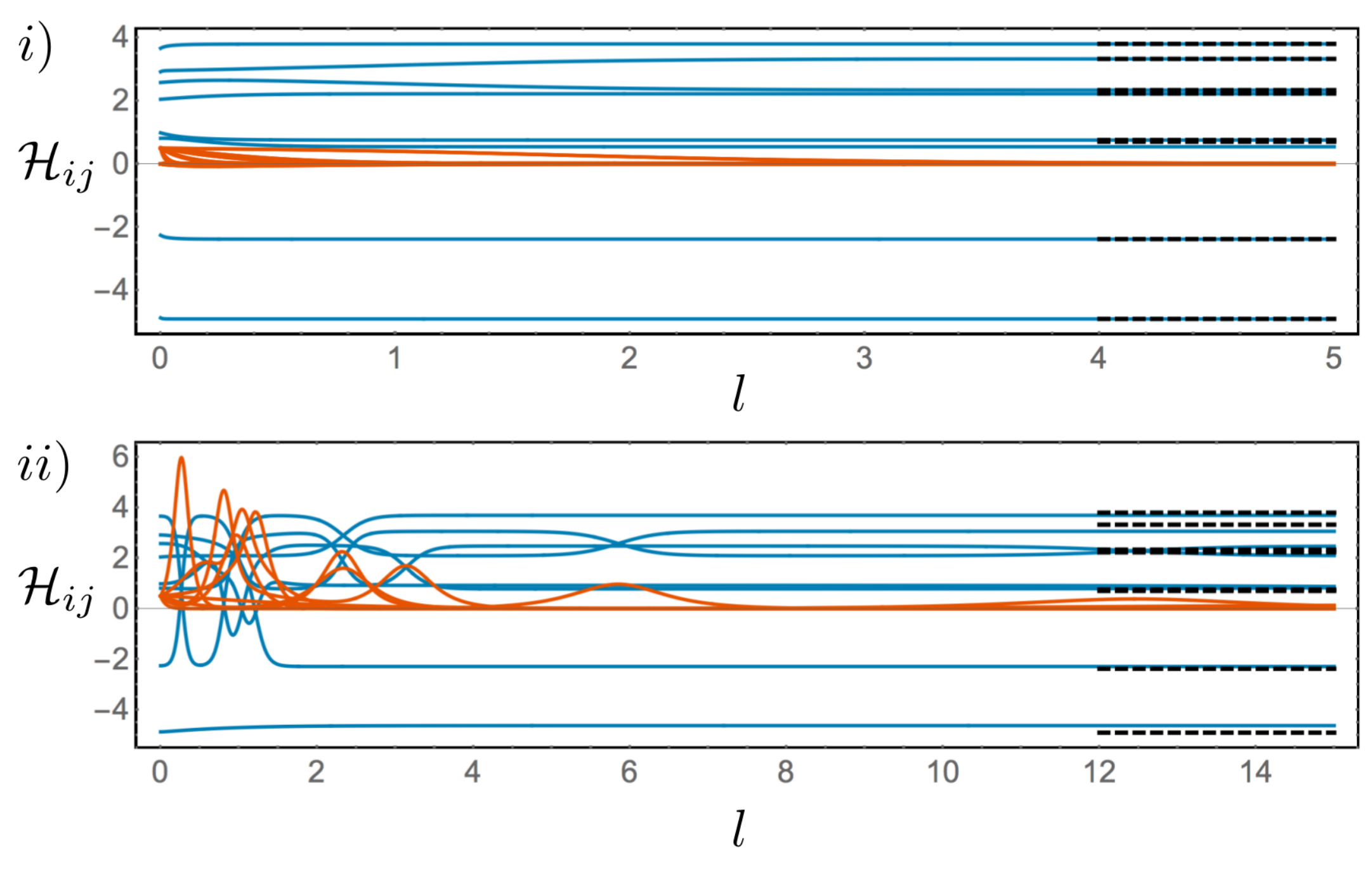}}
\caption{Comparison of i) Wegner and ii) Toda flows for an Anderson insulator of size $L=8$, $W=5$ and $J_0=0.5$. The blue lines indicate the random on-site energies $h_i$, and the orange lines the hopping coefficients $J_{ij}$. The black dashed lines indicate the true eigenvalues of the system: the same disorder realisation is used for both panels. While both methods will eventually converge to the correct eigenvalues, the Wegner flow does so faster and more smoothly. The oscillatory behaviour of the Toda flow is due to the rearrangement of the eigenvalues into ascending order: note that the crossing points of pairs of diagonal elements (blue lines) is accompanied by a temporary increase in the corresponding off-diagonal element (orange lines). This means that accurate numerical integration requires a much finer grid of points in flow time $l$. While Toda flow preserves the sparsity of the Hamiltonian, Wegner flow typically yields better results due to the smoother convergence.} 
\end{figure}\label{fig:flow} 
\subsection{Flow Equations}

The flow equations can be computed from:
\begin{align}
\frac{\ud \mathcal{H}}{\ud l} = [\eta(l),\mathcal{H}(l)],
\end{align}
and the equations for the running couplings read off from the resulting expression. There are $\mathcal{O}(N^2)$ flow equations which must be simultaneously solved\footnote{Symmetries such as $J_{ij} = J_{ji}$ are useful here in reducing the total number of equations.}, and their explicit expressions are given by:
\begin{align}
\frac{\ud J_{ij}}{\ud l} &= -J_{ij} (h_{i} -h_{j} )^{2} - \sum_{k} J_{ik} J_{kj} (2h_k -h_i-h_j), \label{eq.flowJ} \\
\frac{\ud h_i}{\ud l} &= 2 \sum_{j} J_{ij}^{2} (h_i -h_j). \label{eq.flowh} 
\end{align}
The key term is the first term in Eq.~\ref{eq.flowJ} which is always negative and proportional to the square of the on-site energy difference $(h_i-h_j)$. This term guarantees that the off-diagonal elements will decay, and ensures that the terms corresponding to the largest energy splitting decay fastest. This is reminiscent of typical renormalisation group calculations, where high-energy degrees of freedom are eliminated before low-energy ones, and this separation of energy scales gives an appealing hierarchical structure to the flow equation diagonalisation process. However, note that when $h_i \approx h_j$, the corresponding couplings $J_{ij}$ will decay only exponentially slowly: for disorder distributions which are narrow relative to the system size, there will be many such terms and the flow can take a long time to converge. This means that for translationally invariant systems, nothing will flow under the action of this transform: in fact, the Wegner generator evaluates to zero and the system will not be diagonalised. In order to diagonalise a translationally invariant model using this approach, one must either break the translational invariance explicitly (i.e. by setting $h_i \to h_i + \epsilon$ where $\epsilon$ is some small number), or by Fourier transforming into momentum space where the new diagonal elements $\tilde{h}_k$ will no longer all be identical. The flow equations for an initial Hamiltonian formulated in momentum space are identical to Eqs. [\ref{eq.flowJ},\ref{eq.flowh}] but with the site labels swapped for momentum labels. We only note this in passing, however, as for disordered systems such as we consider here there is no real benefit to working in momentum space.
For comparison, we also show the flow equations which can be obtained using the Toda generator. They take a cosmetically simpler form, with only $\mathcal{O}(N)$ equations, each containing no summations over lattice sites:
\begin{align}
\frac{\ud J_{i}}{\ud l} &= 2J_i (h_{i+1}-h_i), \\
\frac{\ud h_i}{\ud l} &= J_i^2-J_{i-1}^2,
\end{align}
however, the flow generated by these expressions turns out to be numerically more challenging to keep track of.

The flow of the variables is shown in Fig.~1 for a small example system of size $L=8$. The eigenvalues are indicated by the dashed black lines. We see that for Wegner flow, the flow is smooth and well-behaved, with the off-diagonal terms $J_{ij}$ all flowing continuously to zero in the large-$l$ limit, and the diagonal terms quickly approach the exact eigenvalues of the model. Toda flow still converges to the correct eigenvalues, but the flow is much more oscillatory. In fact, this problem worsens with increasing system size, due to the much larger number of `swaps' of the diagonal matrix elements. In this particular example, there are two sets of near-dengenerate eigenvalues: careful inspection of Fig.~1 illustrates that, due to these near-degeneracies, neither flow method is quite converged on the timescale shown here. We can obtain the eigenvalues to arbitrary accuracy by integrating the flow equations to arbitrarily long times, however it is clear from Fig.~1 that the flow becomes much slower at longer times. Taking advantage of the slow flow and smooth convergence properties of the Wegner generator, it is convenient to use a logarithmically-spaced grid of points in flow time $l$, reducing the computational cost and allowing efficient access to large maximum flow times.  Due to the numerical difficulty of accurately implementing the Toda generator, we choose to instead focus on the more robust Wegner generator, at the cost of having $\mathcal{O}(N^2)$ variables rather than $\mathcal{O}(N)$.  From this point on, we shall use the Wegner flow exclusively.

\section{Non Interacting Fermions: Power-Law Banded Random Matrix}
As the Wegner flow does not preserve the sparsity of the Hamiltonian, we lose essentially no efficiency in starting from microscopic models which are non-sparse. As a particularly interesting example, we can take the Anderson model with long range, power-law-distributed random hopping 
\begin{align}\label{eqn:HPRBM}
\mathcal{H} = \sum_{ij} J_{ij} c^{\dagger}_i c_{j} + \sum_i  h_i n_i,
\end{align}
where the matrix elements $J_{ij}$ are drawn randomly from a distribution with a width that decays like an inverse power of the distance, i.e. $\sigma(J_{ij}) \propto J_0 / |i-j|^{\alpha}$. The on-site disorder terms $h_i$ are drawn from a box distribution, defined between $[0,W]$, following Ref.~\cite{Quito+16}. The flow equations are the same as for the Anderson model discussed before, see Eq.~(\ref{eq.flowJ},\ref{eq.flowh}), with the only difference being the initial condition for the flow of the hopping matrix.

This model~\ref{eqn:HPRBM} is known as the Power-Law Banded Random Matrix (PRBM) model, and will be the main focus of the rest of this manuscript. Despite being non-interacting, it has an Anderson transition even in $d=1$ as a function of the power-law exponent $\alpha$. Indeed, there is a critical point at $\alpha = d$ separating localised ($\alpha >d$) and delocalised $\alpha<d$) phases respectively \cite{Mirlin+96,Levitov99}. We will not go into full detail here, but precisely at the critical point, the eigenfunctions of the system are multifractal and display unusual critical behaviour that has been the subject of significant scrutiny in Refs. \cite{Varga+00,Mirlin+00,Evers+00,Cuevas+01,Kravtsov+05}. This model has a metal-insulator transition driven by the formation of long-range resonances, where a resonance is defined as the situation when two sites satisfy $J_{ij} > |h_i - h_j|$.  Of particular importance for our purposes here is that though the one-dimensional system exhibits a clear localisation transition at $\alpha=1$ independently of the disorder distributions chosen for the $h_i$ and $J_{ij}$ variables, the statistical properties of the eigenstates vary dramatically for different choices of the power-law exponent $\alpha$. For example, the statistical properties in the region $\alpha < 0.5$ can be described by a Gaussian Orthogonal Ensemble (GOE), while between $0.5 \leq \alpha \leq 1$ the system exhibits intermediate statistics where some properties appear to be GOE but others do not \cite{Mirlin+96}. In the regime $\alpha >1$, the eigenstates are localised but not exponentially: instead, they have power-law tails, which renders their properties somewhat different from typical Anderson localised systems. One may then ask what the consequences of the long-range power-law behaviour are on the dynamical properties and how the dynamics may differ in these different regions: this will be the focus of the present study.

The static properties of this model were studied using flow equations by Ref. \cite{Quito+16}, augmented by a strong-disorder renormalisation group calculation, which showed that a distinct regime with intermediate statistics could be found in the range $0.5 \leq \alpha \leq 1$ by following the flow of the distribution of off-diagonal elements $J_{ij}$, allowing the position of critical region to be obtained directly from the flow rather than from, e.g. eigenstate properties. Our results are consistent with those of Ref. \cite{Quito+16} and so we do not reproduce their work here: instead, we will show some alternative calculations and in particular will focus on how the flow equation method can be used to study the \emph{dynamics} of this model following a quantum quench, even allowing us to compute the dynamical evolution of complex composite objects such as correlation functions.

\subsection{Eigenvalues and level statistics}

Before demonstrating how to compute non-equilibrium dynamical quantities using the flow equation method, let us first show that the method can accurately obtain the static properties of the Hamiltonian, namely the eigenvalues and related quantities. Using the Wegner generator, the eigenvalues converge rapidly\footnote{The precise convergence depends on each individual disorder realisation and system size, but eigenvalues can typically be determined to good precision at modest flow times on the order of $l_{max} \sim 10^{1}\textrm{-}10^{2}$. To compute the level spacing ratio, however, we require much greater precision and a larger $l_{max}$.}, as shown in Fig. 1, and can be easily shown to match the exact eigenvalues of the model to arbitrary numerical precision depending on the choice of $l_{max}$. Instead of demonstrating the convergence of the eigenvalues, here we focus on computing a more demanding quantity. Analagously to studies of many-body localisation which proposed the investigation of the statistics of the many-body energy levels \cite{Oganesyan+07}, by computing the level statistics of the eigenvalues obtained via the flow equation method, we can extract the single-particle level-spacing statistics by computing the following quantities:
\begin{align}
\delta_n &= |E_n - E_{n+1}|, \\
r_n &= \textrm{min}(\delta_n,\delta_{n+1})/\textrm{max}(\delta_n,\delta_{n+1}),
\end{align}
In a delocalised phase which exhibits level repulsion, the levels should be distributed according to the Wigner-Dyson probability distribution $P(\delta) = (\pi/2) \delta \exp(\pi \delta^2 /2)$, leading to an average value of the level spacing ratio $r \approx 0.53$ (where the averaged value is denoted $r \equiv \overline{\langle r_n \rangle}$, and is first averaged within a sample, then again over disorder realisations) in accordance with random matrix theory \cite{Kravtsov09}. In localised phases where randomness dominates and there is no level repulsion, the level spacings will be distributed according to a Poisson distribution $P(\delta) = \exp(-\delta)$, leading to $r \approx 0.39$. There is no mobility edge in this system and so we do not restrict our averages to the centre of the spectrum, as is common in studies of many-body localisation.
\begin{figure}
\resizebox{\linewidth}{!}{\includegraphics{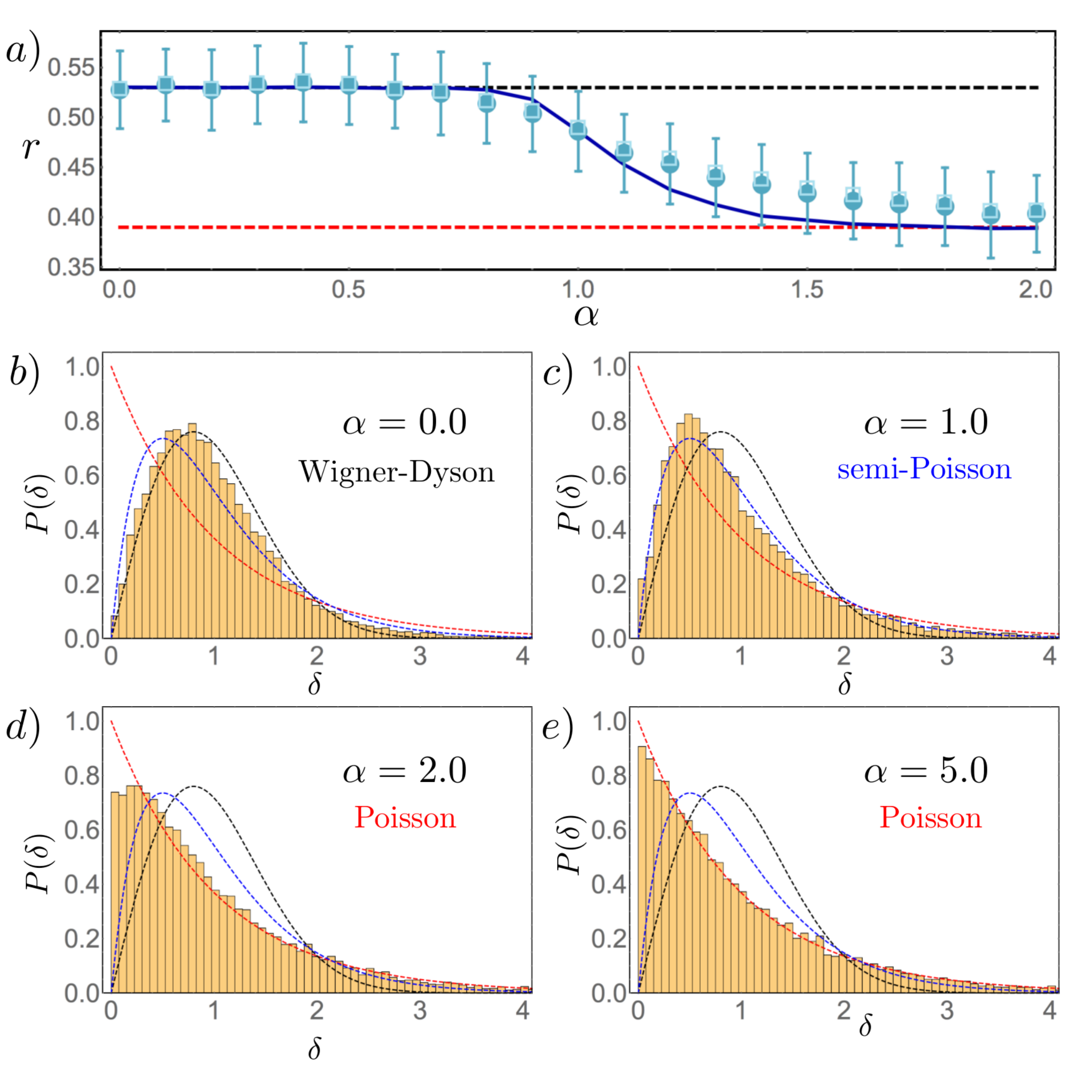}}
\label{fig:level_spacing} 
\caption{Level spacing statistics for chains of length $L=64$, averaged over $N_s = 256$ disorder realisations. a) A comparison of the mean level spacing ratio $r$, averaged over samples and disorder realisations, computed by exactly diagonalising the Hamiltonian (light blue empty squares) and with flow equations (dark blue filled circles, with error bars showing the standard deviation across disorder realisations: on this scale, the variance is smaller than the plot markers). There is excellent agreement between the flow equation and exact results for all values of the power-law exponent. The solid blue line shows the mean level spacing ratio computed exactly for a system of size $L=4096$ and averaged over $50$ disorder realisations, illustrating the finite-size effects particularly present in the region $0.5\leq \alpha \leq 1.5$. The black and red dashed lines show the Wigner-Dyson (GOE) and Poisson values of $r$ respectively. Panels b-d) show that the shape of the full distribution of level spacings $\delta_n$ for a variety of values of the exponent $\alpha$, illustrating how it changes from a Wigner-Dyson distribution (black dashed line) at small $\alpha$ to a Poisson distribution (red dashed line) at large $\alpha$, and resembles a semi-Poisson distribution (blue dashed line) at the critical point $\alpha_c = 1.0$.} 
\end{figure}
This is a particularly demanding quantity to compute because it is sensitive to exponentially small energy level differences, and because (as shown in Fig. 1) the flow equations methods only asymptotically approach the true diagonal Hamiltonian. While it is possible to obtain the eigenvalues to several significant figures in only a modest flow time, much higher accuracy is required in order to extract the averaged level spacing ratio $r$, which requires both long flow times and accurate numerical integration routines. Being able to compute this quantity accurately is therefore an excellent test of the flow equation method.

For the PRBM model which dislays a phase transition, we should expect to see the averaged value of $r$ change across the transition, shown in Fig. \ref{fig:1}. At small values of the power-law exponent $\alpha \leq 0.5$, the system tends towards Wigner-Dyson level statistics, signifying delocalisation, whereas for $\alpha>1.5$ the level statistics converge to the Poisson distribution. Histograms of the full level spacing distribution $P(\delta)$ are also shown in Fig. 1, demonstrating the excellent agreement with Poisson statistics at large $\alpha$, Wigner-Dyson statistics at small $\alpha$ and resembles intermediate semi-Poisson statistics in the critical region close to the transition, where $P(\delta) = 4 \delta \exp(-2 \delta)$, though with small deviations \cite{Varga+00}. Note that the region $0.5 < \alpha < 1.5$ is particularly vulnerable to to finite-size effects, as the thermodynamic limit is approached only very slowly in this model \cite{Varga+00}. Previous studies have shown \cite{Mirlin+96} that in the region $0.5 < \alpha < 1$, the system is is an intermediate configuration where certain quantities may appear to follow the Gaussian Orthogonal Ensemble (GOE) values but the overall statistics do not in fact conform to this distribution.

\subsection{Dynamics}
We can compute the dynamics of any operator by transforming the operator into the basis which diagonalises the Hamiltonian, time-evolving with respect to the diagonal Hamiltonian, and then flowing the operator back into the physical basis. We demonstrate this with the number operator $n_i(t)$. Since the Hamiltonian is quadratic, the flow of this operator at scale $l$ takes a closed form, which reads:
\begin{align}
n_i(l,t=0) &= \sum_{j} \alpha^{(i)}_j(l) n_j + \sum_{jk} \beta^{(i)}_{jk}(l) c^{\dagger}_j c_k \label{eq.nansatz}
\end{align}
where the coefficients are now explicit functions of the fictitious flow time $l$.
The flow equations for this operator can be obtained by computing $n_i'(l) = [\eta(l), n_i(l)]$ using the same $\eta(l)$ which diagonalises the Hamiltonian and are given by:
\begin{align}
\frac{\ud \alpha^{(i)}_j}{\ud l} &= -2 \sum_k J_{jk} (h_k - h_j) \beta_{kj}^{(i)}, \\
\frac{\ud \beta_{jk}^{(i)}}{\ud l} &= -J_{jk} (h_k-h_j)(\alpha^{(i)}_k-\alpha^{(i)}_j) \nn\\
&-\sum_{n} \left[ J_{nj} (h_n - h_j) \beta^{(i)}_{nk} + J_{nk} (h_n - h_k) \beta^{(i)}_{nj} \right], \label{eq.nflow}
\end{align}
 
After transforming $n_i(t=0)$ in the new basis, we can time-evolve it with respect to the diagonal Hamiltonian $\tilde{\mathcal{H}} = \sum_k \epsilon_k \tilde{n}_k$ by solving the Heisenberg equation of motion:
\begin{align}
i \frac{\ud n_i}{\ud t} &= \sum_{jk} \beta_{jk} (\tilde{h}_k - \tilde{h}_j) c^{\dagger}_j c_k
\end{align}
such that the time-evolved operator in the diagonal basis is given by:
\begin{align}
\tilde{n}_i(l = \infty, t) &=  \sum_{j} \alpha^{(i)}_j(l) n_j + \sum_{jk} \beta^{(i)}_{jk}(l) \textrm{e}^{-i (\tilde{h}_k - \tilde{h}_j) t} c^{\dagger}_j c_k
\end{align}
We then use the flow equations (Eqs. \ref{eq.nflow}) to transform the number operator back into the original basis, where it will take the form:
\begin{align}
n_i(l=0,t) &= \sum_{j} \alpha^{(i)}_j(t) n_j + \sum_{jk} \beta^{(i)}_{jk}(t) c^{\dagger}_j c_k
\end{align}
which looks the same as Eq.~\ref{eq.nansatz}, but now back in the original $l=0$ basis of the microscopic Hamiltonian and with \emph{time-evolved} coefficients $\alpha^{(i)}_j(t)$ and $\beta^{(i)}_{jk}(t)$. In computing the `backwards' evolution from the $l=\infty$ basis to the original $l=0$ basis, it is extremely useful to store the flow of $\mathcal{H}(l)$ in the memory, as the $l=\infty$ initial conditions for the $J_{ij}(l)$ terms are exponentially small and attempting to re-compute $\eta(l)$ on the fly will result in significant numerical errors. This `backwards' flow must be re-computed each time step we wish to consider. However, because the time evolution step in the diagonal basis is essentially just multiplication by a phase, we can immediately compute the long-time behaviour of $n_i(t)$ without having to compute all intermediate timesteps. 

When calculating the dynamics following a quench from an initial product state, as we will do here, the average $\langle n_i (t) \rangle$ reduces to:
\begin{align}
\langle n_i (t) \rangle = \sum_{j} \alpha^{(i)}_j(t) \langle n_j \rangle
\end{align}
where the average is computed with respect to the initial product state. In principal, we can choose any starting state including equilibrium states of the model, however here we shall restrict ourselves to quenches starting from an initial charge density wave product state, i.e. $\ket{\psi} = \ket{010101...}$. Using the formalism described in the previous section, we can compute the dynamics following a quench from such a N\'eel state. 

By computing the time-evolved density on each lattice site, we can construct the \emph{imbalance} following a quench from a N\'eel state. This is essentially the analogue of a `staggered magnetisation' for spinless fermions, and is defined as:
\begin{align}
\mathcal{I}(t) = \frac{2}{N} \sum_i (-1)^{i} \langle n_i (t) \rangle 
\end{align}
In the delocalised state, this imbalance should quickly decay to zero, whereas in the localised state it should remain finite. This is no longer a local quantity, and requires computing the occupation of every single lattice site at each timestep. The results of this compuation are shown in Fig. 3, where we see that it behaves as expected. There is no clear sign of a transition, however there is a broad crossover region which matches the same feature in the level statistics.

\begin{figure}
\resizebox{0.99\linewidth}{!}{\includegraphics{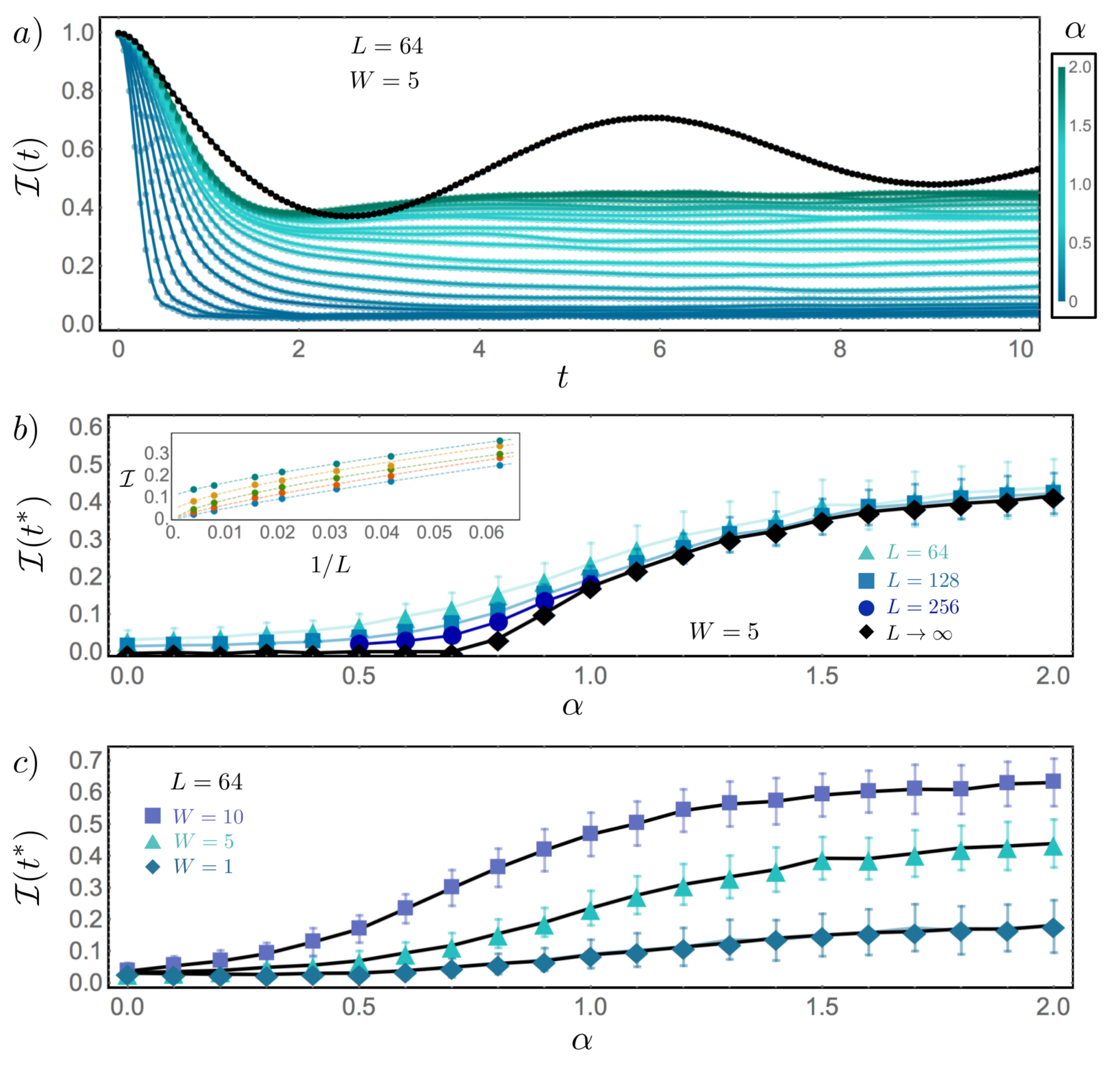}}
\label{fig:imb} 
\caption{a) The evolution of the imbalance for a chain of length $L=64$ following a quench from a N\'eel state, for various values of the power-law exponent $\alpha$ from $0.0$ (bottom) to $2.0$ (top) in increments of $0.1$, averaged over $\sim200$ disorder realisations. The black curve is the Anderson insulator, effectively the $\alpha \to \infty$ limit, for comparison. b) The endpoint $\mathcal{I}(t^{*})$ of the imbalance following a quench, plotted at $t^{*}=100$ as a function of the power-law exponent $\alpha$ for three different system sizes: $L=64$ (triangles), $L=128$ (squares), $L=256$ (circles). There is a broad continuous crossover between localised and delocalised phases. The black line is the result of extrapolation to the thermodynamic limit: a linear fit in the inverse system size (with $L=16,24,32,64,128,256$) is used everywhere apart from the region $0.5 \leq \alpha <1$ where finite-size effects are strongest and the fit is power-law in the inverse system size, as shown in the inset. Due to the strong dependence on system size, it is very likely that this naive procedure still overestimates the imbalance in the critical region near $\alpha \approx 1$.  c) The imbalance computed for fixed system size $L=64$ and for four different values of the on-site disorder strength $W=1.0,5.0,10.0$ (from bottom to top). While the position of the critical point is unaffected by our choice of $W$, the imbalance varies dramatically, making the transition more or less visible from the dynamics as we vary the disorder strength but also increasing the strength of the finite-size effects which lead to a finite imbalance for $\alpha < 1$.} 
\end{figure}

In the intermediate critical region $0.5 \leq \alpha \leq 1$, there is a persistent imbalance at long times for all system sizes we have investigated thus far, though this imbalance does weakly decrease with increasing system size. This is not simply due to slow dynamics in this region: we have checked up to times $t \sim 10^3$ hopping times and seen no decay of this imbalance. We have performed a simple finite-size scaling analysis: for all values of $\alpha$ outwith the region $0.5 \leq \alpha < 1$, we have used a linear fit in the inverse system size to extrapolate the imbalance in the thermodynamic limit. Within this region, where finite-size effects are the strongest, we have used a power-law fit in the inverse system size (shown in the inset of Fig. 2b) to attempt to capture the scaling towards the thermodynamic limit. As shown in the inset, the dependence of the imbalance on $1/L$ becomes increasingly non-linear as $L \to \infty$, though as $\alpha \to 1$ this non-linearity moves to larger and larger system sizes. This is consistent with the idea that the number of resonances which lead to delocalisation scale like $L^{1-\alpha}$ in this region \cite{Quito+16}: to be able to see full delocalisation (and a zero imbalance), one must go to larger and larger system sizes in the limit of $\alpha \to 1$. Therefore, while naive finite-size scaling presents an improvement over the raw data, it likely still overestimates the imbalance in the critical region due to the limits of the finite system sizes we have simulated. The imbalance does not saturate immediately at the critical point of $\alpha = 1$, and continues to grow slowly with increasing $\alpha$. This is due to the eigenstates in this region having long power-law tails \cite{Yeung+87} : while they may be formally localised, these long tails still facilitate some degree of transport, and only in the $\alpha \to \infty$ limit when the tails are sufficiently small is the transport completely inhibited on all length scales.

%\textcolor{red}{MS:I would remove this paragraph about bad metal phase. }
%The other possibility is that there may exist a so-called `bad metal' phase within the critical region $0.5 < \alpha <1$, resulting in a finite imbalance that would persist in the $L \to \infty$ limit: while this cannot be excluded based on our data, we have carefully examined the relaxation properties of $\mathcal{I}(t)$ in this region and find no signs of the power-law relaxation that one would expect to see in this case. This `bad metal' region, where the eigenstates are delocalised but the dynamics retain a quasi-localised nature, is likely a finite-size effect, but one which is important for current-generation experiments which are limited to system sizes where this behaviour is very likely to make an appearance.

The final value of the imbalance also depends strongly on the strength of the on-site disorder. This is again consistent with the picture that the delocalisation is caused by long-range resonances, i.e. events where $J_{ij} > |h_i - h_j|$. The narrower the distribution of the on-site disorder, the smaller the typical value of the energy difference $|h_i - h_j|$ and the more likely it is that delocalising resonances will be observed. Similarly, at large disorder the resonances will be suppressed, necessitating larger system sizes in order to see the true behaviour of the system in the thermodynamic limit.

  It is known from previous work that despite the deocalised nature of all eigenstates in this region, fluctuations around the GOE values are large \cite{Mirlin+96}: a further, more detailed investigation of the dynamics in this region is required in order to umambiguously determine the behaviour of the imbalance and other dynamical quantities in the thermodynamic limit.

\begin{figure*}
\resizebox{\linewidth}{!}{\includegraphics{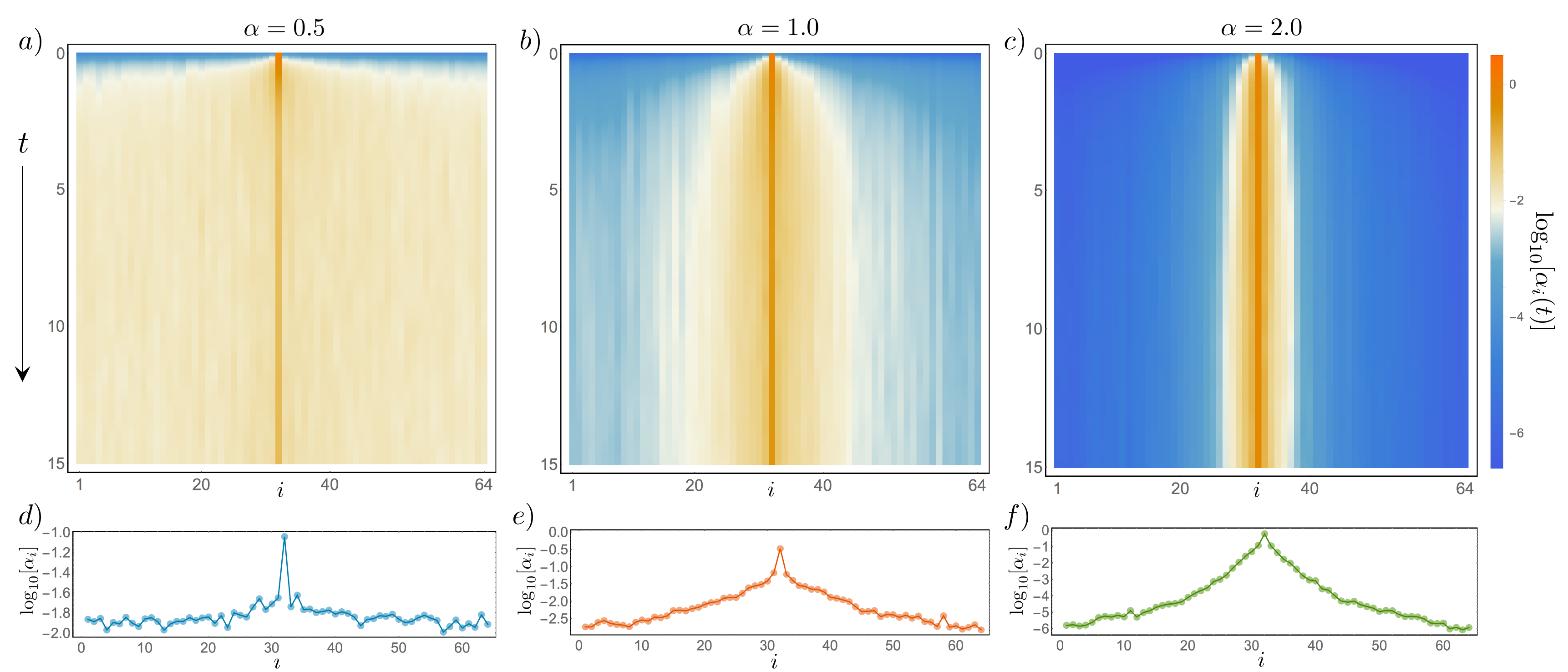}}
\label{fig.ns} 
\caption{The real-time spreading of the density operator initially defined on the central lattice site for a system of size $L=64$, on-site disorder strength $W=5$ and $J_0 = 0.5$. Panels a-c) show a density plot of the logarithm of the support of the density operator in both space and time for three values of the exponent $\alpha=0.5,1.0,2.0$ corresponding to the delocalised, the critical and the localised regions respectively. Panels d-f) show the support of the density operator at a fixed time $t^{*}=100$ following a quench from the initial CDW state, for the same values of the exponents $\alpha$. In the delocalised phase, there is a peak on the lattice site where the density operator is initially defined, followed by a broad tail that is roughly constant with distance, due to the long-range coupling. In the localised regions, the same initial peak is present, but there are long, slowly decaying tails. Note that the tails do not decay exponentially, as would be the case for an Anderson insulator, but decay much more slowly.} 
\end{figure*}
\subsection{Operator spreading}

We can also ask how the support of an operator initially defined on a single lattice site varies in \emph{real time}. Taking the number operator $n_i(t)$ again, we can compute the diagonal terms $\alpha^{(i)}_j(t)$ and see how the operator spreads across the lattice as time goes on. By writing operators in this way, the flow equation method provides a natural framework in which to calculate and understand how operator spreading takes place even in many-body quantum systems: at $t=0$, the operator in this representation is simply a delta function, with $\alpha^{(i)}_j(t=0) = \delta_{ij}$ and $\beta^{(i)}_{jk} = 0 \phantom{.} \forall j,k$. As the operator evolves under the PRBM Hamiltonian, however, diagonal terms with $j \neq i$ start to become non-zero, and (if the power-law exponent is small enough) eventually the operator will develop non-zero weight across the entire system. Note that this quantity is \emph{independent} of the initial state, as no expectation value is considered here: instead, we are directly accessing a property of the ensemble of Hamiltonians that comprise the PRBM model.

The results of this procedure are shown in Fig. 4 for three different values of the exponent $\alpha$. Although tp the best of our knowledge there are no rigourous Lieb-Robinson-type bounds for light cones in systems with both disorder and power-law couplings (but see for example here~\cite{MinhArxiv19} for a recent work), our results for the operator spreading are consistent with other numerical results in very similar systems for the dynamical evolution of probability distributions \cite{Roy+18} and out-of-time-ordered correlation functions \cite{Luitz+19,Chen+19}. For small $\alpha$, the results are consistent with other works indicating that the number operator initially spreads in time like $t \sim \log(r)$ (where $r$ represents the width of the support of the operator), with full delocalisation occuring on very short timescales. For larger $\alpha$ on the other hand, there is an initial expansion of the number operator at short times before the full effects of localisation take hold and the support of the operator remains confined to a narrow region. For intermediate values of $\alpha$, the behaviour of the operator is less clear: the combination of strong disorder and random long-range hoppings leads to complex dynamics with many competing timescales and no clear `light cone'. 

\subsection{Correlation functions}

One great advantage of the flow equation approach is that at the end of the procedure, we have all the information about the transformed operator $n_i(t)$ independently of any state. This allows us to easily compute quench dynamics with respect to different initial states, for example, but also allows us to build more complex correlation functions at essentially no extra computational cost. As an example, here we will define and compute the equal-time density-density correlation function, although we could also examine higher-order correlation functions. The connected correlation function is defined as:
\begin{align}
C_{ij}^{(c)} &= \langle n_i(t) n_j (t) \rangle - \langle n_i(t) \rangle \langle n_j (t) \rangle \\
&= \sum_{kq} \beta_{kq}^{(i)}(t) \beta_{qk}^{(j)}(t) \langle n_k \rangle (1- \langle n_q \rangle) 
\end{align}
where the averages are computed with respect to an initial product state, and any averages of the form $\langle c^{\dagger}_i c_j \rangle$ are zero for $i \neq j$. Defining the density-density correlation function with respect to the central site of the chain as:
\begin{align}
C_j(t) &= \langle n_{L/2}(t) n_j(t) \rangle - \langle n_{L/2}(t)\rangle \langle n_j(t) \rangle
\end{align}
we can follow the dynamical evolution of this correlator starting from an initial N\'eel state. The results can be plotted in a similar manner to Fig. 4 and the results are qualitatively the same, so we do not plot them here but instead focus on the long-time behaviour.

\begin{figure}
\resizebox{0.99\linewidth}{!}{\includegraphics{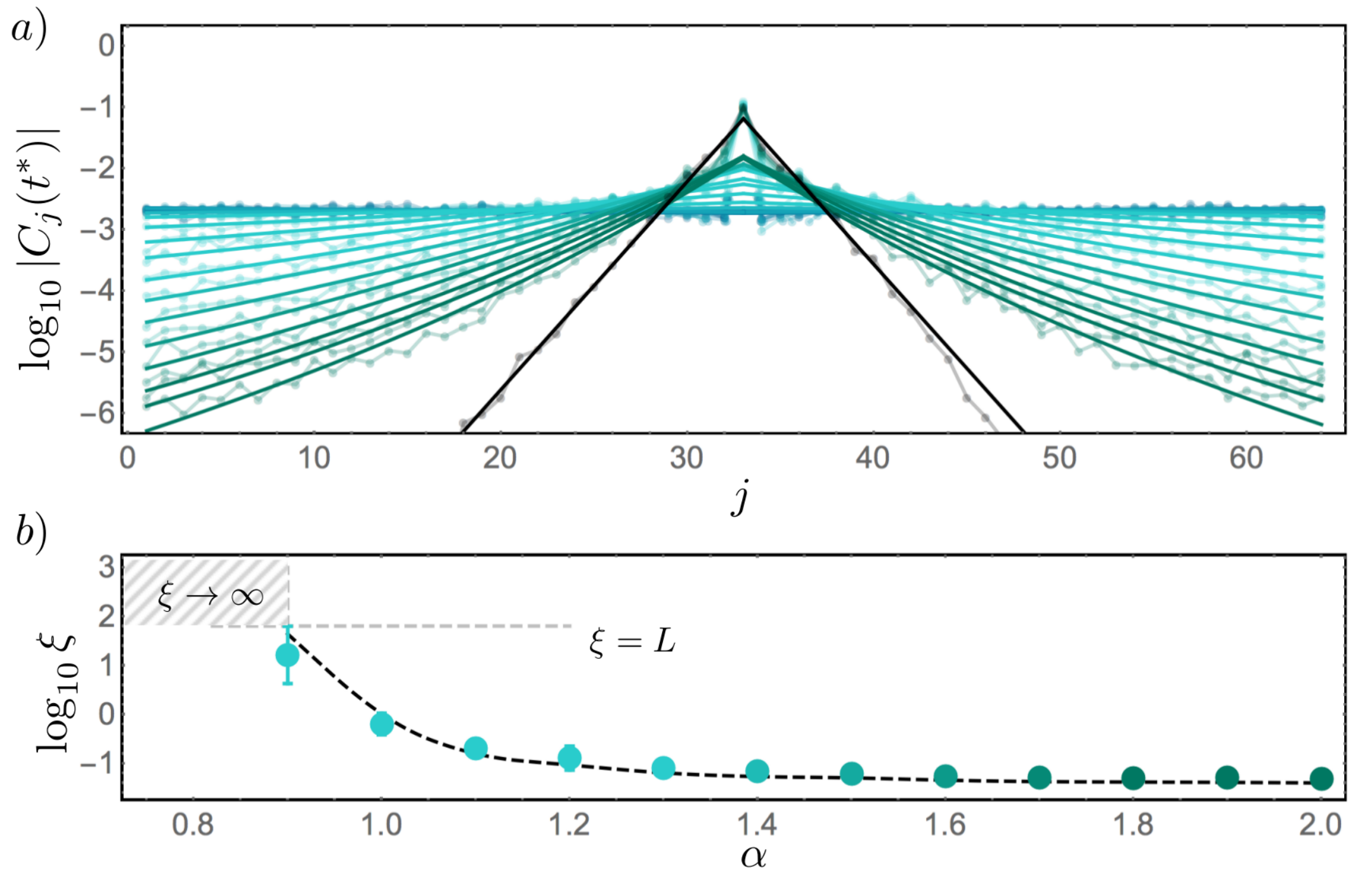}}
\label{fig:1} 
\caption{a) The density-density correlation function measured with respect to the central site, computed at time $t^{*}=100$ after a quench for a chain of length $L=64$, $W=5$ and $J_0=0.5$ for various values of the power-law exponent $\alpha$ from $0.0$ (top) to $2.0$ (bottom) and averaged over $256$ disorder realisations. The colour scale is the same as in Fig. 3. The points joined by translucent lines are the raw data points, while the solid lines show the fits used to extract the correlation length. (Note that the fits are performed with respect to the long distance tails, not to the central region.) At small $\alpha$, the correlation function is essentially flat. At large $\alpha$, there is an initial exponential decay away from the centre, followed by a long power-law tail. The larger $\alpha$ is, the larger the region of exponential decay before the power-law tail beging. The black line shows the correlation function of the Anderson insulator (the $\alpha \to \infty$ limit) for comparison: it is exponentially localised, and decays quickly away from the central site. b) The behaviour of the correlation length $\xi$ extracted from the fits shown in panel a) above. The error bars represent the uncertainties in the fit: they are smaller than the plot markers for most values of $\alpha$. For $\alpha \geq 1$, $\xi$ decays monotonically. At $\alpha = 0.9$, the correlation length is of the same order as the system size, and below this value the correlation function is flat and the correlation length is essentially infinite.} 
\end{figure}

We can compute the long-time limit of this correlation function $C_j(t^{*})$, shown in Fig. 5, to examine the steady-state behaviour reached long after a quench. For large values of $\alpha$, the correlation function displays an initial exponential decay followed by long, slow power-law tails, mirroring the known behaviour of the eigenstates themselves \cite{Yeung+87}, as well as the spreading of the density operator as discussed in the previous section. For small $\alpha$, the correlation function rapidly becomes flat, signifying the existence of long-range correlations in the delocalised phases.

A correlation length can be extracted by fitting the long-distance tails of $C_{j}(t^*)$ to a decaying functional form. The fits are performed for a power-law decrease with distance for $\alpha \geq 1$, e.g. $C_r \sim (r/\xi)^{\gamma}$. Additionally, in order to obtain the best statistics possible, the fits were performed on symmetrised data where the correlation function was averaged about the central point. The raw data for the correlation functions, the symmetrised fits and the correlation length extracted from them can be found in Fig. 5. For $\alpha < 0.9$, the correlation function is essentially flat on the length scales considered here, meaning that the correlation length is larger than the system size and cannot be accurately fitted with either power-law or exponential form, suggesting that the system is delocalised at these values of the exponent $\alpha$, consistent with the known phase diagram of this model. Note that this correlation length may not be the same as the localisation length of the system: rather, it is simply a lengthscale that can be associated with the decay of the (dynamical) correlation functions a long time after a quench.

\section{Interacting Quantum Systems}
\label{sec:int}

So far in this work we have limited our attention to non-interacting but disordered quantum systems, for which the application of flow equation method is particularly transparent as it can be carried out without further approximations. Of course, a large variety of other methods already exist which can treat non-interacting systems exactly, so let us now highlight some other applications of the flow equation method to systems which are extremely difficult for other numerical methods. In presence of interactions and disorder, such as in models relevant for many-body localization (MBL), the application of the method becomes more challenging, due to the exponential growth of the Hilbert space or to the generation of higher-order couplings along the flow. Yet, this is exactly the situation in which, due to the challenging nature of the problem and the lack of general methodologies, developments of such techniques could result in interesting applications.

One line of attack which has been explored in the literature is to formulate the flow equation in the full Hilbert space, as an exact procedure to diagonalize the Hamiltonian~(\cite{Pekker+17}) or to formulate the flow of operators in the language of tensor networks taking advantage of the power of matrix product state representation~\cite{Yu+19}. The latter is a particularly promising numerically exact implementation of the flow equation method, allowing simulation of larger system sizes than, e.g. exact diagonalisation. Extension of this method to compute dynamical quantities, i.e. using a generalisation of the time-dependent variational principle, would be an extremely interesting avenue of future investigation.

An alternative point of view, which we have explored recently in the context of MBL~\cite{Thomson+18} is to obtain an effective Hamiltonian by truncating the flow of higher order couplings, taking advantage from the fact that within the MBL phase the effective $l$-bit model describes interactions which decay exponentially as a function of distance and the higher-order terms in this fixed-point Hamiltonian therefore contribute exponentially less weight at each term \cite{Imbrie16a}. We refer the reader to our previous work for the full details \cite{Thomson+18}, but to briefly illustrate how this method works, consider the case of an interacting fermionic system given by the Hamiltonian:
\begin{align}
\mathcal{H}(l=0) = \sum_{i} J c^{\dagger}_i c_{i+1} + \sum_i h_i n_i + \sum_{i} \Delta_{i} n_i n_{i+1}.
\end{align}
Under the action of the unitary flow generated by the Wegner generator acting in real space, this Hamiltonian will still be brought into a diagonal form, but higher-order interaction terms will be successively generated, leading to a fixed-point Hamiltonian of the form:
\begin{align}
\tilde{\mathcal{H}}(l=\infty) = \sum_i h_i n_i + \frac12 \sum_{ij} \Delta_{ij} n_i n_j + \sum_{ijk} \Gamma_{ijk} n_i n_j n_k + ...
\end{align}
containing arbitrarily high-order terms. In strongly disordered systems, these new high-order terms are exponentially suppressed, and thus even the most extreme approximation of truncating the Hamiltonian at order $\Delta$ is accurate deep in the localised phase. This truncation can also be made in the case where the interactions are weak and the newly-generated higher-order terms will be negligible.

While this truncation naturally represents an approximation, we can quantify the accuracy of this procedure by noting that unitary transforms conserve certain quantities, particularly traces of powers of the Hamiltonian $\textrm{Tr}[\mathcal{H}^p]$ (where $p \geq 1$ is an integer). These quantities are known as invariants of the flow. If the invariant computed with the initial $l=0$ Hamiltonian differs from the one computed using the $l \to \infty$ Hamiltonian obtained by the flow procedure, that means that the truncation has introduced an error: computing this diference this gives us a way to measure the accuracy of the method such that it can be applied with confidence.

In the flow equation scheme, truncating at the leading order in the interactions means that the interacting system can be treated at the same computational cost as the non-interacting system, namely $\mathcal{O}(N^2)$, albeit with a large prefactor due to the increased number of terms in the running Hamiltonian. This is a considerably better scaling than the $\mathcal{O}(2^N)$ size of the Hilbert space, demonstrating the power of this method for suitable interacting quantum systems. This avenue of investigation is precisely where we believe the true promise of the flow equation method lies.
There are, however, several complications which one must be aware of in attempting to apply this method to interacting systems, the most severe of which is the necessity to adopt a consistent ordering scheme for the operators while calculating high-order commutators. The resulting normal-ordering procedure, detailed in \cite{Kehrein07}, leads to the generation of `correction terms' which include feedback from higher-order terms in the running Hamiltonian. This is a powerful approach which can ultimately turn the flow equation method into a sophisticated non-perturbative tool \cite{Kehrein07}, however in combination with the tendency of this method to generate all possible allowed couplings, it can lead to a high level of algebraic complexity that limits extension of the method to higher-order interactions: work to alleviate this problem is currently ongoing, and for further technical details see Refs. \cite{Kehrein07,Thomson+18,Kelly+19}.

Extension of this method to disordered, interacting systems with long-range couplings and disordered, interacting Floquet systems (e.g. time crystals) is currently ongoing \cite{Thomson+19a,Thomson+19b}. It is also possible to extend the Toda generator to cover interacting spin chains, as discussed in detail in Ref. \cite{Monthus16}, however we have not investigated this possibility numerically due to the difficulty in implementing Toda flow for the simpler case of non-interacting systems.

\section{Conclusion}
\label{sec:end}

We have presented in this manuscript a study of the quench dynamics of the Power-Law Random Banded Matrix model, a system of non-interacting fermions with long-range hopping that is often used to study features of the Anderson transition and critical point. We have shown that despite the eigenstates being formally delocalised for power-law exponents $\alpha<1$, the non-equilibrium dynamics starting from a charge density wave initial state retain a quasi-localised nature in the region $0.5 \leq \alpha <1$ that persists in the imbalance up to the largest system sizes we are currently able to simulate in reasonable computational time using the flow equation method. We have further characterized the evolution of operator spreading across the localization transition and the long-time, real-space, profile of density-density correlations.

There are many open directions for future development of the flow equation framework as applied to disordered, many-body quantum systems. Firstly, only a very small number of generators have been studied in the literature, with the Wegner generator (and modifications of it) being the most commonly-encountered choice. While this generator is numerically robust, it is not necessarily the optimum choice, particularly for \emph{interacting} systems: further investigation into appropriate generators of unitary transforms would be desirable. Secondly, throughout the formalism presented here we have not investigated the properties of wavefunctions themselves, choosing instead to focus on computing local quantities accessible to experiments, however as quantities such as the inverse participation ratio and entanglement entropy require knowledge of the wavefunctions, and as far-from-equilibrium phenomena such as MBL can involve eigenstate properties, it would be desirable and presumably straightforward to extend the method to compute eigenstates as well as eigenvalues. Thirdly, the computational aspects of flow equations as applied to many-body systems have not historically received a great deal of attention (with notable exceptions, e.g. \cite{Savitz+17}): further developments and optimisations in coding and applying continuous unitary transforms could pave the way to a more widespread adoption of this method.

{\bf Author contributions - } The calculations and numerical simulations were performed by SJT. The manuscript was jointly written by SJT and MS.

\bibliographystyle{unsrt}
\bibliography{refs}

\end{document}